\documentclass[twocolumn,eqsecnum,superscriptaddress,showpacs]{revtex4}
\usepackage{epsfig}
\begin{document}
\title{ NUCLEON EDM FROM ATOMIC SYSTEMS AND CONSTRAINTS ON 
SUPERSYMMETRY 
PARAMETERS  }

\author{Sachiko Oshima}
\affiliation{Department of Physics, Faculty of Science, Tokyo Institute 
of Technology, Tokyo, Japan}
\author{Takeshi Nihei}\email{nihei@phys.cst.nihon-u.ac.jp}
\affiliation{Department of Physics, Faculty of Science and Technology, 
Nihon University, Tokyo, Japan}
\author{Takehisa Fujita}\email{fffujita@phys.cst.nihon-u.ac.jp}
\affiliation{Department of Physics, Faculty of Science and Technology, 
Nihon University, Tokyo, Japan}

\date{\today}%

\begin{abstract}

The nucleon EDM is shown to be directly related to the EDM of 
atomic systems. From the observed EDM values of the atomic Hg system, 
the neutron EDM can be extracted, which gives a very stringent 
constraint on the supersymmetry parameters. It is also shown that 
the measurement of Nitrogen and Thallium atomic systems should provide 
important information on the flavor dependence of the quark EDM.  
We perform numerical analyses on the EDM of neutron, proton and 
electron in the minimal supersymmetric standard model with
CP-violating phases. 
We demonstrate that the new limit on the neutron EDM extracted from 
atomic systems excludes a wide parameter region of supersymmetry breaking 
masses above 1 TeV, 
while the old limit excludes only a small mass region below 1 TeV. 
\end{abstract}


\maketitle

\vspace{1cm}
\noindent

\newpage
\section{\bf Introduction}

In field theory, it is important to understand the property of the symmetry 
of the Lagrangian density. In the standard model, the time reversal invariance 
is normally kept, but due to the phase of Kobayashi-Maskawa mixing matrix, 
there is the T-violation term present in the Lagrangian density. The phase 
shows up as the CP violation effects on the $K^0-{\bar K}^0$ mixing 
in the kaon decay processes \cite{bel,bab}. 
Since the CP violation of the Kaon decay experiment 
determines the Kobayashi-Maskawa phase, it should be interesting to see 
what should be further physical observables which are predicted from the CP violation 
phenomena or T-violating observables. 

Recent experiments on the CP violation of the $B^0-{\bar B}^0$ mixing 
in the B-meson decay processes suggest that the phase of Kobayashi-Maskawa 
mixing matrix may not be able to describe the observed mixing angle from 
the $B^0-{\bar B}^0$ mixing. 
This suggests that there may be some other terms present in the Lagrangian 
density, which violates the T-invariance. 

The direct measurement of the T-violation effects can be examined if 
one can measure the electric dipole moments (EDM) of the isolated system. 
The finite number of the neutron EDM should exhibit most clearly the direct 
evidence of the T-violation effects in the Lagrangian density. 
Until now, the upper limit of the neutron EDM $d_n$ from the direct measurements 
is around \cite{ne1, ne2, q1}
$$ d_n \simeq (1.9\pm 5.4)\times 10^{-26} \  {\rm e}\cdot {\rm cm}  \eqno{(1.1a)} $$ 
$$ d_n \simeq (2.6\pm 4.0 \pm 1.6)\times 10^{-26} \  {\rm e}\cdot {\rm cm} . 
\eqno{(1.1b)} $$ 
This EDM is still not finite, but gives a strong constraint on the supersymmetry 
model parameters. 

There are also EDM  measurements of atomic systems such as
$^{129}{\rm Xe}$ \cite{w22,w2} and $^{199}{\rm Hg}$ \cite{q22,q2}.  However, 
it is not very easy to extract the individual particles EDM from the atomic 
EDM measurements since there is the Schiff theorem \cite{q3}. 
The Schiff theorem states that the EDM of the individual particles cannot 
be measured if the constituents are interacting with nonrelativistic 
static Coulomb force. The EDM of the individual particles should show up 
when they are interacting with relativistic kinematics \cite{t1,t2,q44} or with strong 
interactions \cite{q13}. The EDM due to the former case is found 
in the heavy atomic system 
while the EDM of the latter case arises from the finite nuclear size effects. 

Recent careful studies clarify \cite{edm1} that the EDM arising 
from the nuclear finite size effects is directly related to the neutron EDM $d_n$ as 
$$ d_{\rm Xe} \simeq  1.6 d_n \eqno{(1.2a)} $$
$$ d_{\rm Hg} \simeq -2.8 d_n \eqno{(1.2b)} $$
where $ d_{\rm Xe}$ and $ d_{\rm Hg}$ denote the EDMs of Xe and Hg atomic systems. 
In particular, the EDM of $^{199}$Hg atomic system is observed with high accuracy and 
from the observed vaule of $ d_{\rm Hg}$, we can deduce the neutron EDM as
$$ d_n \simeq (0.37\pm 0.17 \pm 0.14)\times 10^{-28} \  {\rm e}\cdot 
{\rm cm} . \eqno{(1.3)} $$ 
This number is three orders of magnitude smaller than the direct measurement of 
neutron EDM in eq.(1.1). Therefore, it should be worthwhile to examine 
how this number of the neutron EDM can give constraints on the supersymmetry model 
parameters. 

Supersymmetry (SUSY) is one of the most promising candidates for new physics 
beyond the standard model \cite{review-susy}. 
In particular, supersymmetric models contain new CP-violating phases
\cite{gknos} 
in addition to the phase of the Cabbibo-Kobayashi-Maskawa mixing matrix, 
and the effect of these phases may explain the discrepancy between 
the experimental result and the standard model prediction 
for CP-asymmetry in the B-meson decays. 
In general, the new CP-violating phases in supersymmetric models 
induce a non-vanishing EDM for quarks and leptons. 
In fact, they give a large contribution to the EDMs at the one-loop level. 
On the other hand, it is known that the standard model contributions 
to the EDMs cancel out up to the two-loop level, 
and the leading three-loop contribution is much smaller than the current 
experimental upper bound. 
This implies that a large EDM is not only a signal 
of CP-violation but also an indirect evidence of SUSY.

In the present work, we calculate the EDM of neutron, proton and 
electron in the minimal supersymmetric standard model (MSSM) 
with CP-violating phases. 
We shall show that the neutron EDM limit in eq.(1.3) gives a very severe 
constraint on the parameters in the supersymmetric model. 
For a sizable value of the CP-violating phases, 
the magnitude of soft breaking mass parameters typically has to be larger
than several TeV to satisfy the EDM constraint.

Further, we show, for the first time, that the proton EDM can be extracted 
from the Nitrogen and Thallium atomic EDM. This is simply because, 
for the $^{15}{\rm N}$ case, it has 
the ground state with spin $1/2$, and the state is described by the proton hole 
state as $ |\pi (1p_{1\over 2})^{-1}:{1\over 2}^- \rangle $ to a good accuracy. 
Also, for the $^{205}{\rm T{\ell}}$ case, since it is a single proton hole state, 
it can be expressed as 
$ |\pi (3s_{1\over 2})^{-1}:{1\over 2}^+ \rangle $ to a good approximation 
even though the neutrons are not in the magic but in the two hole state. 
 
From the experimental constraint, the nucleus should have the spin $1/2$ 
such that the quadrupole field should not have any effects on the EDM measurements. 
In this respect, $^{15}N$ and  $^{205}{\rm T{\ell}}$ atomic systems  
must be good cases for measuring the nuclear EDM. 
The EDM of the Nitrogen atomic system $d_{^{15}{\rm N}} $ 
and the $^{205}{\rm T{\ell}}$ atomic system $ d_{\rm T{\ell}}$ 
are related to the proton EDM $d_p$, and the calculated numbers become 
$$ d_{^{15}{\rm N}} \simeq - 0.16 d_p  \eqno{(1.4a)} $$
$$ d_{\rm T{\ell}} \simeq 4.8 d_p . \eqno{(1.4b)} $$

Recently, Regan et.al have measured the atomic EDM of $^{205}{\rm T{\ell}}$ 
by the atomic beam magnetic resonance method \cite{reg}. From the measurement 
of the atomic EDM of $^{205}{\rm T{\ell}}$, we can extract the proton EDM $d_p$
$$ d_p \simeq -(0.83\pm 0.90) \times 
 10^{-25} \  {\rm e}\cdot {\rm cm}  \eqno{(1.5)} $$
which is quite a severe constraint on the proton EDM $d_p$. 
Here, we have not included the electron EDM contribution to the atomic EDM of 
$^{205}{\rm T{\ell}}$, and the electron EDM contribution will be discussed 
in a qualitative manner in section 4. 

On the other hand, the EDM for the Nitrogen case is smaller than the ones 
in heavier nuclei,  but we believe that the Nitrogen EDM should be also measured in 
future.

This paper is organized as follows. In the next section, we briefly discuss 
the new mechanism to obtain the EDM of the individual particles in nucleus, 
like neutron EDM or proton EDM from atomic systems. 
Then, in section 3, we evaluate the EDM of neutron, proton and 
electron in the minimal supersymmetric standard model 
with CP-violating phases, and find various constraints on the supersymmetric 
parameters from the observed neutron EDM value. 
In section 4, we present 
a qualitative discussion on the contributions of the electron  
and the proton EDMs to the atomic EDM value.  
Section 5 summarizes 
what we clarify the nuclear EDM and 
the related parameters which appear in the supersymmetry models.

\vspace{0.5cm} 

\section{\bf Nucleon EDM from atomic systems }

The electric dipole moment (EDM) of the isolated system can present 
a direct evidence of the T-violation interaction if it is finite. 
Since the measurement is carried out for the total atomic system, 
the EDM of the individual particles (electron EDM or nucleon EDM) 
should be extracted from the atomic EDM measurement.   
However, this extraction is a nontrivial task since there is the Schiff theorem. 
The Schiff theorem states that the EDM of the individual particles cannot be measured 
if the neutral system is interacting with nonrelativistic electrostatic forces. 

The EDM of electron becomes important when the system has a relativistic 
effect since the Schiff theorem is not applicable to the relativistic case.  
Indeed for heavier atoms, the relativistic effect becomes important since 
the relativistic correction is seen as $(Z\alpha)^2$. 
For the electron EDM, there are many calculations
in heavy atoms. 

If the interactions are due to nuclear forces, then there is no effect 
from the Schiff theorem. Since the nucleus is always found in atomic system, 
the nuclear EDM can be seen as the finite size effects in the total 
atomic system. Recent calculations on the nuclear EDM show that there is 
an appreciably large effect of the nuclear EDM to the total atomic EDM after 
the effects due to the Schiff theorem are taken into account. 
The results obtained in \cite{edm1} for Xe and Hg atomic EDM are summarized 
in eqs. (1.2).

Now, we briefly describe the finite size effects of the atomic EDM since 
the detailed calculation is found in  \cite{edm1}.  In particular, the effects 
due to the Schiff theorem are well treated and explained in  \cite{edm1}, and therefore 
we do not repeat it for this part in this paper.

\subsection{ Hamiltonian of atomic systems} 

We first write the Hamiltonian 
of the total atomic and nuclear systems for Nitrogen. 

The unperturbed Hamiltonian $H_0$ of the Nitrogen system can be written
$$ H_0 =  \sum_{i=1}^Z \left[ { {{\bf p}_i}^2\over{2m} } -
\sum_{j=1}^Z{e^2\over{|{\bf r}_i-{\bf R}_j| }} \right] +
{1\over 2}\sum_{i\not= j}^Z{e^2\over{|{\bf r}_i-{\bf r}_j| }} $$ 
$$ +\sum_{i=1}^A{ {\bf P}_i^2\over{2M} }+{1\over 2}\sum_{i\not= j}^A 
V_{NN}(|{\bf R}_i-{\bf R}_j|)+
{1\over 2}\sum_{i \not= j}^Z{e^2\over{|{\bf R}_i-{\bf R}_j| }}  \eqno{(2.1)} $$
where ${\bf r}_i$, ${\bf p}_i$ denote the coordinate and the momentum 
of the electron while   ${\bf R}_i$, ${\bf P}_i$ denote the nuclear variable 
and momentum, respectively. 

On the other hand, the perturbed Hamiltonian coming from the EDM is written as 
$$ H_{edm}= -  \sum_{i=1}^Z\sum_{j=1}^Z {e{\bf d}_e^i \cdot ({\bf r}_i-{\bf R}_j )
\over{|{\bf r}_i-{\bf R}_j|^3}} 
+\sum_{i=1}^Z\sum_{j\not= i}^Z {e{\bf d}_e^i \cdot ({\bf r}_i-{\bf r}_j )
\over{|{\bf r}_i-{\bf r}_j|^3}}  $$
$$ -\sum_{i=1}^Z\sum_{j=1}^A {e{\bf d}_N^j \cdot ({\bf r}_i-{\bf R}_j )
\over{|{\bf r}_i-{\bf R}_j|^3}}  
-\sum_{i=1}^A\sum_{j\not= i}^Z {e{\bf d}_N^i \cdot ({\bf R}_i-{\bf R}_j )
\over{|{\bf R}_i-{\bf R}_j|^3}}  $$
$$-\sum_{i=1}^Z{\bf d}_e^i\cdot {\bf E}_{ext} 
-\sum_{i=1}^A{\bf d}_N^i\cdot {\bf E}_{ext}
+e\sum_{i=1}^Z ({\bf r}_i-{\bf R}_i )\cdot {\bf E}_{ext}
 \eqno{(2.2)} $$
where the summation over $Z$ in nucleus means that it should be taken 
over protons. The EDM of the nucleon can be expressed in terms of the 
nucleon isospin as 
$$ {\bf d}_N^i = {1\over 2}\left[(1+\tau_i^z)d_p \mbox{\boldmath $\sigma$}^i +(1-\tau_i^z) 
d_n\mbox{\boldmath $\sigma$}^i  \right] .  \eqno{(2.3)} $$

\subsection{ Nuclear EDM from nuclear excitation} 

Here, we evaluate the finite size effects on the second order EDM 
energy in the nucleus. The perturbed EDM Hamiltonian for nuclear parts 
$H_{edm}^{(n)}$ can be written as
$$ H_{edm}^{(n)} =  -\sum_{i=1}^A\sum_{j\not= i}^Z e{\bf d}_N^i \cdot  {({\bf R}_i-{\bf R}_j )
\over{|{\bf R}_i-{\bf R}_j|^3}}  
-{e\over 2} \sum_{i=1}^A (1+\tau_i^z) {\bf R}_i\cdot {\bf E}_{ext} . 
  \eqno{(2.4)} $$

Now, we consider the second order EDM energy due to the intermediate 
nuclear excitations, keeping the atomic state in the ground state. 
This process arises from the finite nuclear size effects in the EDM Hamiltonian. 
The second order EDM energy can be written as
$$ \Delta E^{(2)}_{fs} =-\sum_n {e^2\over{E_n-E_0 }}
\langle \Psi_{N}|\sum_{i=1}^A\tau_i^z {\bf R}_i\cdot {\bf E}_{ext}
  |n \rangle  $$
$$ \times \langle n|  \sum_{i\not= j}^A 
 {1\over 4}\left[(1+\tau_i^z)d_p \mbox{\boldmath $\sigma$}^i +(1-\tau_i^z) 
d_n \mbox{\boldmath $\sigma$}^i  \right]   $$
$$ \cdot{(1+\tau_j^z)\left( {\bf R}_i-{\bf R}_j \right)
\over{|{\bf R}_i-{\bf R}_j|^3}} 
|\Psi_{N} \rangle \eqno{(2.5)} $$
where $E_0$ and $E_n$ denote the ground state energy and the excitation energy 
of the nuclear states, respectively. 
The electron states are kept in the ground state throughout the calculation, and 
therefore it is not written here. Also, it should be noted that 
we made use of the relation  $\displaystyle{ \sum_{i=1}^A {\bf R}_i =0 }$ 
since we set the center mass coordinate to the nuclear center.

\subsection{Nitrogen atomic system}

Now, we calculate the $^{15}{\rm N}$ case in which we assume a simple 
single particle shell model state $ |\pi (1p_{1\over 2})^{-1} \rangle $. This  
should be rather a good description for $^{15}{\rm N}$. 
Further, the atomic states stay in the ground state, 
and therefore the electron wave functions are not written here.  
Thus, we write the nuclear wave function as 
$$ |\Psi_{N}\rangle =  |\pi (1p_{1\over 2})^{-1}: {1\over 2}^{-} \rangle  . \eqno{(2.6)} $$ 
In this case, the intermediate states $|n \rangle $ that contribute to the second 
order EDM energy [eq.(2.6)] for $\pi (1p_{1\over 2}) $ are restricted to 
$$ |n\rangle = |\pi (1s_{1\over 2})^{-1} :{1\over 2}^{+} \rangle ,\quad 
|\pi (1p_{1\over 2})^{-2}, \pi (2s_{1\over 2}):{1\over 2}^{+}  \rangle . \eqno{(2.7)} $$ 
With eqs.(2.5-7), we carry out numerical calculations of the second order EDM energy 
$ \Delta E^{(2)}_{fs} $ and obtain the relation between the Nitrogen EDM $d_{^{15}{\rm N}}
$ 
and the proton EDM $d_p$ as 
$$ d_{^{15}{\rm N}} \simeq -0.16 d_p . \eqno{(1.4a)} $$
Even though the factor in front of $d_p$ is not very large number, we believe 
that the EDM of Nitrogen can be well measured experimentally in future.

\subsection{Thallium atomic system}

Now, we calculate the $^{205}{\rm T{\ell}}$ case in which we assume a simple 
single particle shell model state $ |\pi (3s_{1\over 2})^{-1} \rangle $. This 
should be reasonable for the $^{205}{\rm T{\ell}}$ case. 
Therefore, the nuclear wave function for  $^{205}{\rm T{\ell}}$ can be written as
$$ |\Psi_{N}\rangle =  |\pi (3s_{1\over 2})^{-1}: {1\over 2}^{+} \rangle .  
\eqno{(2.8)} $$ 
In this case, the intermediate states $|n \rangle $ that contribute to the second 
order EDM energy [eq.(2.6)] for $| \pi (3s_{1\over 2})^{-1}\rangle  $ are restricted to 
$$ |n\rangle = |\pi (2p_{1\over 2})^{-1} :{1\over 2}^{-} \rangle ,\quad 
|\pi (3s_{1\over 2})^{-2}, \pi (3p_{1\over 2}):{1\over 2}^{-}  \rangle . \eqno{(2.9)} $$ 
With eqs.(2.8-9), we carry out numerical calculations of the second order EDM energy 
$ \Delta E^{(2)}_{fs} $ and obtain the relation between the Thallium EDM $d_{\rm T{\ell}}$ 
and the proton EDM $d_p$ as 
$$ d_{\rm T{\ell}} \simeq 4.8 d_p . \eqno{(1.4b)} $$
Clearly, the EDM of Thallium is quite large, and therefore there must be 
a good chance of observing it. 
Indeed, there is a measurement of T$\ell$F compound system, and 
Cho el al.  \cite{thal} obtained the EDM of $d_{\rm T\ell F} $ as
$$ d_{\rm T\ell F} \simeq (-1.7\pm 2.9) \times 
 10^{-23} \  {\rm e}\cdot {\rm cm} .  \eqno{(2.10)} $$
Making use of the relation between Thallium EDM and proton EDM [eq.(1.4b)], 
we obtain the proton EDM as
$$ d_p \simeq (-0.35\pm 0.60) \times 
 10^{-23} \  {\rm e}\cdot {\rm cm}  \eqno{(2.11)} $$
where we ignored the contribution of EDM from the Fluoride atomic system. 

Further, Regan et al. have recently measured the atomic EDM of $^{205}{\rm T{\ell}}$ 
by the atomic beam magnetic resonance method \cite{reg}. They extracted the electron 
EDM from their measurement, and obtained the constraint of electron EDM $d_e$. 
Here, we estimate the proton EDM from their measurements, and the atomic EDM of 
$^{205}{\rm T{\ell}}$ is found to be
$$ d_{\rm T\ell } \simeq -(0.4\pm 0.43) \times 
 10^{-24} \  {\rm e}\cdot {\rm cm} .   \eqno{(2.12)} $$
Therefore, we can extract the proton EDM using eq.(1.4b)
$$ d_p \simeq -(0.83\pm 0.90) \times 
 10^{-25} \  {\rm e}\cdot {\rm cm}  \eqno{(2.13)} $$
which gives stronger constraints than eq.(2.11) of Cho et.al  
almost by two orders of magnitude. This proton value of $d_p$ is just comparable 
to the neutron EDM $d_n$ from the direct measurement. 

As mentioned above, the effects due to the electron EDM $d_e$ in heavy atoms 
should be large. In fact the contribution of the electron EDM to the Thallium case 
is evaluated in \cite{liu}, and they found a very large enhancement factor.  
In this paper, however, we have not included the electron EDM. 
Instead, we will discuss the electron EDM contribution to the atomic EDM 
in a qualitative fashion in section 4.

\vspace{0.5cm} 
\section{Constraints from neutron EDM on supersymmetry model}


In this section, we discuss the constraints from the neutron EDM 
in the MSSM. 
In the present analysis, we work in a framework of a phenomenological MSSM. 
The relevant parameters in this model are defined as follows. 
SUSY breaking mass terms for the squarks and the sleptons contain
a common scalar mass parameter $m_0$ as 
\begin{eqnarray}
-m_0^2 (|\tilde{q}|^2 + |\tilde{u}|^2 + |\tilde{d}|^2 
                      + |\tilde{\ell}|^2 + |\tilde{e}|^2 ), 
\label{eq:scalar_mass}
\end{eqnarray}
where $\tilde{q}$ and $\tilde{\ell}$ denote the SU(2) doublet fields for 
the left-handed squarks and sleptons, respectively, while $\tilde{u}$, 
$\tilde{d}$ and $\tilde{e}$ denote the SU(2) singlet fields for the 
right-handed scalar up-quark, the scalar down-quark and the scalar electron, 
respectively. 
Gaugino mass terms for the bino $\tilde{B}$, the wino $\tilde{W}^a$ ($a=1,2,3$)
and the gluino $\tilde{g}^{\alpha}$ ($\alpha=1,\cdots,8$) 
are given by 
\begin{eqnarray}
-\frac{1}{2} (M_1 \tilde{B}\tilde{B} + M_2 \tilde{W}^a\tilde{W}^a
                         + M_3 \tilde{g}^{\alpha} \tilde{g}^{\alpha})
   + {\rm h.c.}, 
\label{eq:gaugino_mass}
\end{eqnarray}
where $M_1$, $M_2$ and $M_3$ denote
a mass parameter for $U(1)_Y$, $SU(2)_L$ and $SU(3)_C$ gaugino, respectively. 
We assume the GUT relation for the gaugino masses 
$M_1$ $=$ $\frac{5}{3}\tan^2\theta_W M_2$ and 
$M_3$ $=$ $\frac{\alpha_3}{\alpha_2} M_2$.

The SUSY conserving part of the Lagrangian can be written by the 
superpotential 
\begin{eqnarray}
W & = & Y_u \hat{U}^c \hat{H}_2 \hat{Q}  
      + Y_d \hat{D}^c \hat{H}_1 \hat{Q}
      + Y_e \hat{E}^c \hat{H}_1 \hat{L} + \mu \hat{H}_1 \hat{H}_2, 
          \nonumber \\
  &  & 
\label{eq:superpotential}
\end{eqnarray}
where $Y_u$, $Y_d$ and $Y_e$ are Yukawa couplings for the up-quark, 
the down-quark and the electron, respectively. The coefficient $\mu$
is a Higgs mixing mass parameter. 
The symbols with a hat denote chiral superfields with self-evident notations. 
In eq.(\ref{eq:superpotential}), we neglect generation mixings, 
since we discuss generation conserving effects. 
The scalar trilinear couplings are parameterized by the universal 
mass parameter $A$ as 
\begin{eqnarray}
A (Y_u \tilde{u}^* H_2 \tilde{q} + Y_d \tilde{d}^* H_1 \tilde{q}
  + Y_e \tilde{e}^* H_1 \tilde{\ell}). 
\label{eq:aterm}
\end{eqnarray}

Furthermore, the Higgs sector in this model includes the ratio of the vacuum 
expectation values of the two neutral Higgs fields, 
$\tan\beta$ $=$ $\langle H_2 \rangle/\langle H_1 \rangle$,
and a mass parameter $m_A$ for the pseudoscalar neutral Higgs field.

In summary, the Lagrangian is 
parameterized by the following parameters 
\begin{eqnarray}
\tan\beta, \ m_A, \ M_2, \ m_0, \ \mu, \ A. 
\label{eq:parameters}
\end{eqnarray}
Among these, $M_2$, $\mu$ and $A$ can have a non-vanishing complex phase.
We take a phase convention that $M_2$ is real and positive. 
In order to examine the effects of CP-violation, 
we define two complex phases ($\theta_\mu$, $\theta_A$) by 
\begin{eqnarray}
\mu = |\mu|{\rm exp}(i\theta_\mu), \ \ A = |A|{\rm exp}(i\theta_A). 
\end{eqnarray}


The EDMs of quarks and leptons are defined as a coefficient in 
the electric dipole operator 
\begin{eqnarray}
{\cal L} = -\frac{i}{2} d_f \overline{f} \sigma_{\mu\nu}\gamma_5 f F^{\mu\nu}. 
\label{eq:edm_operator}
\end{eqnarray}
The EDM $d_f$ of the fermion $f$ ($=$ $u$, $d$, $e$) can be obtained
by evaluating relevant Feynman diagrams. 
In the case of the MSSM, three kinds of diagrams contribute 
at the one-loop level:
(i) chargino exchange contribution, (ii) neutralino exchange contribution
and (iii) gluino exchange contribution. 
Note that the gluino contribution does not exist for the electron EDM $d_e$. 
The expression of $d_f$ for each fermion can be found in Ref.~\cite{q11}.

To obtain the EDM of the neutron and the proton, we employ 
the non-relativistic quark model as 
\begin{eqnarray}
d_n & = & \frac{1}{3}(4d_d - d_u), \ \ d_p \ = \ \frac{1}{3}(4d_u - d_d). 
\label{eq:dndp}
\end{eqnarray}
It should be noted that a quark chromomagnetic dipole operator and 
a pure gluonic dipole operator may give comparable contributions
to the electric dipole operator (\ref{eq:edm_operator}) \cite{ibrahim-nath}. 
Inclusion of these operators may cause cancellations to suppress the 
neutron EDM. However, this kind of cancellations occurs in relatively small portions of
the whole parameter space. Therefore, we neglect the effects of these 
operators for simplicity. 
Also, we focus on a nuclear finite size effects, and neglect contributions 
from the chromomagnetic dipole operator for the strange quark 
\cite{strange-cdm} in this analysis. 


In the following, we present the results of our numerical analysis. 
In all the figures, 
we always choose $\tan\beta=10$ and $m_A$ $=$ $700\,{\rm GeV}$
as reference values. 
For large $\tan\beta$, the EDMs are typically enhanced so that 
the constraint on the mass parameters becomes more stringent. 
The EDMs are not very sensitive to $m_A$, since the Higgs masses are 
not directly related with the relevant diagrams.

\begin{figure}[p]
\begin{center}
\begin{minipage}{17cm}
\epsfig{file=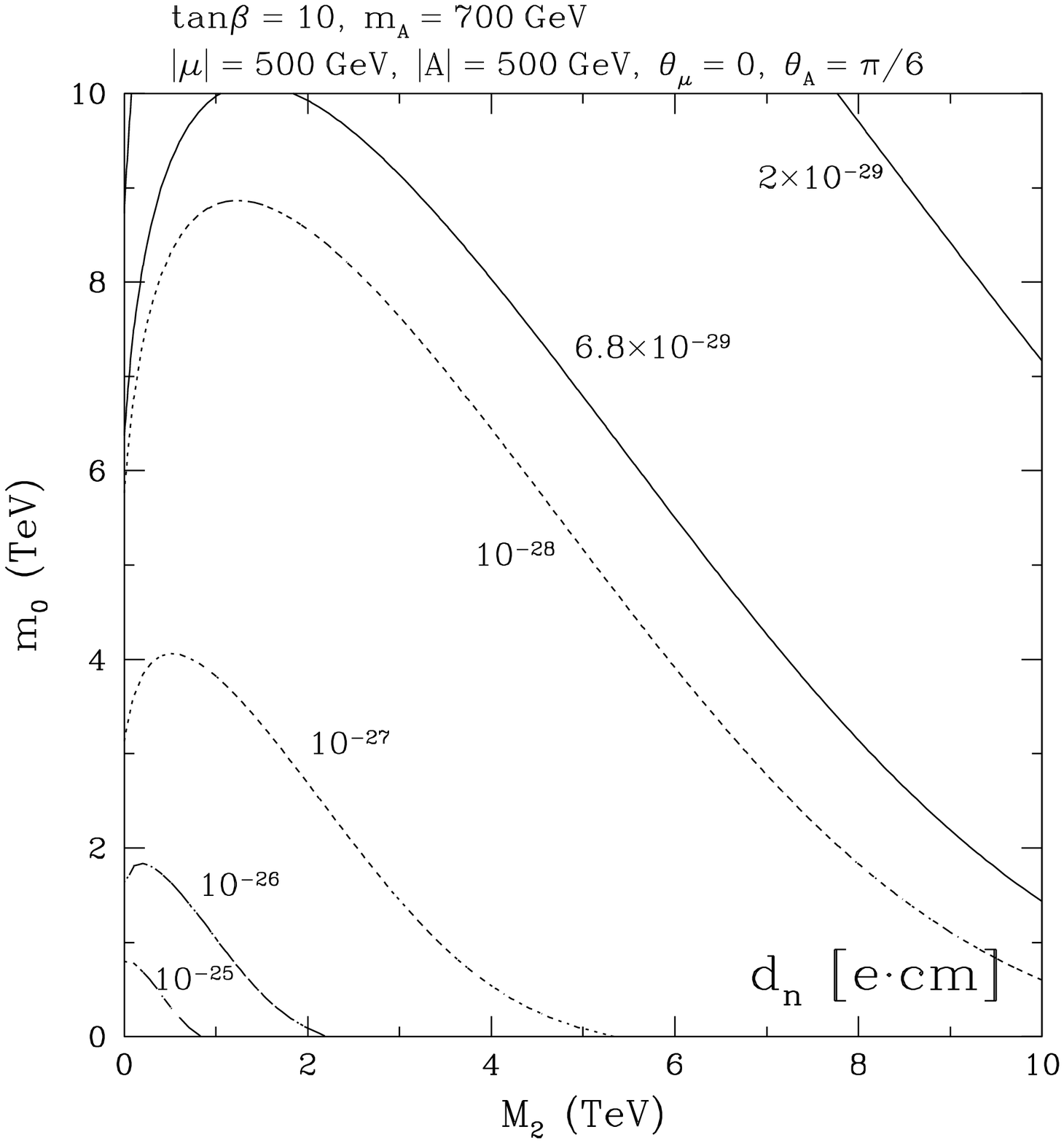,width=8cm} 
\hspace*{1mm}
\epsfig{file=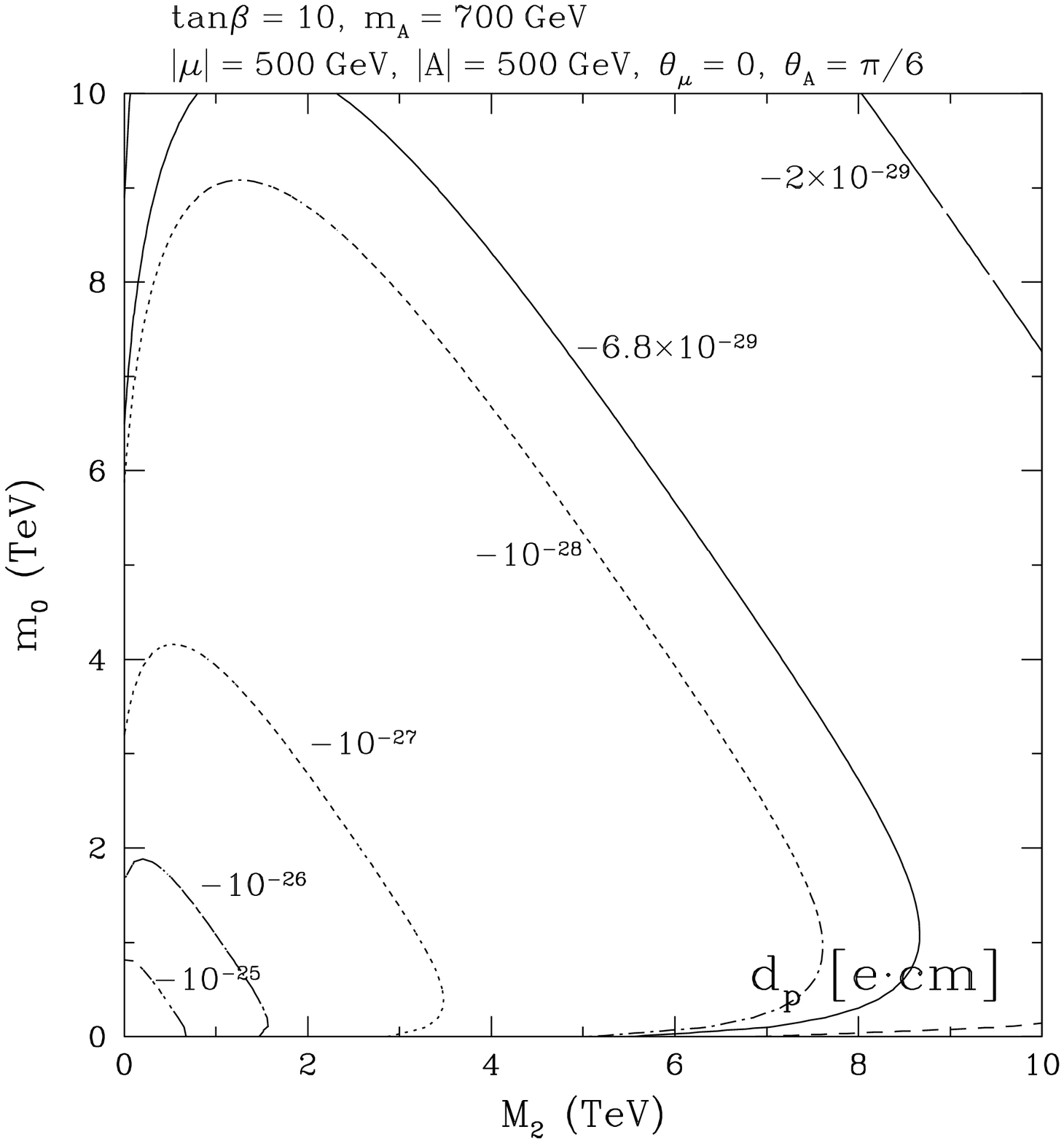,width=8cm} 
\end{minipage}
\begin{minipage}{17cm}
\epsfig{file=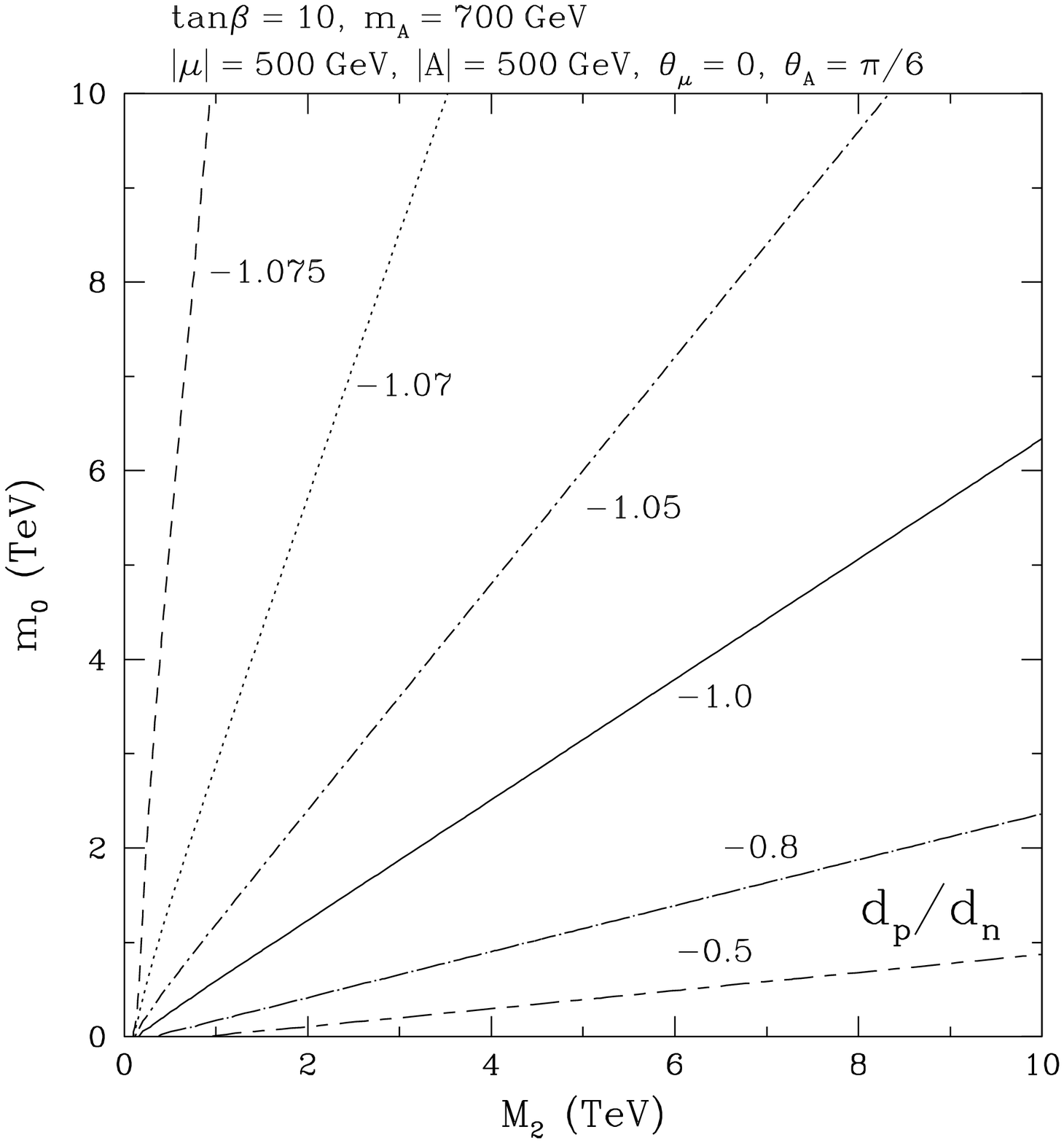,width=8cm} 
\hspace*{1mm}
\epsfig{file=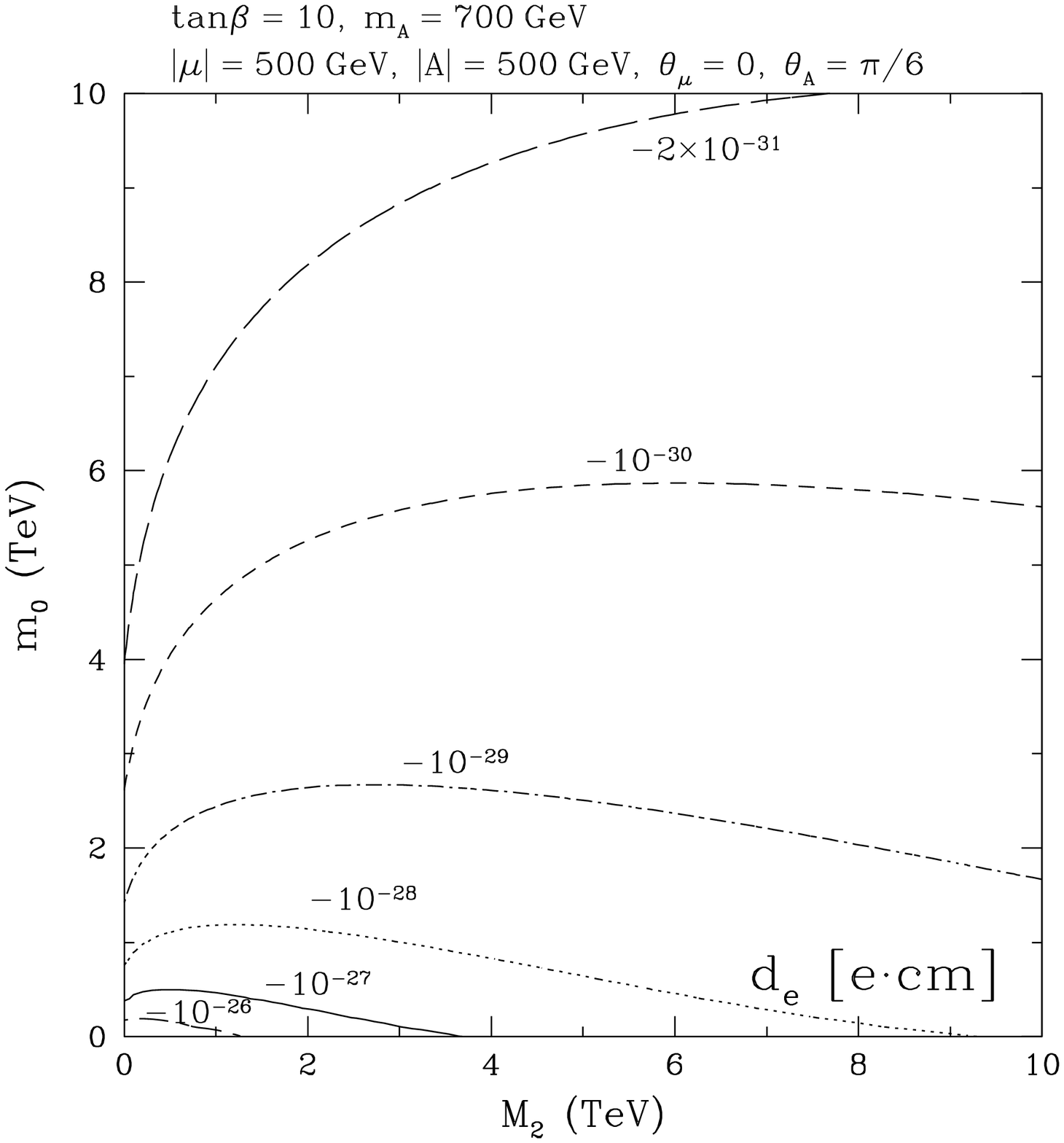,width=8cm} 
\end{minipage}
\end{center}
\caption{\label{fig:cont-m0-m2} 
Contour plots of $d_n$ (upper left window), $d_p$ (upper right window), 
$d_p/d_n$ (lower left window) and $d_e$ (lower right window) 
in the ($M_2$,$m_0$) plane for 
$\tan\beta=10$, $m_A=700\,{\rm GeV}$, 
$|\mu|=500\,{\rm GeV}$, $|A|=500\,{\rm GeV}$, 
$\theta_\mu=0$, $\theta_A=\pi/6$. 
}
\end{figure}

Let us begin with the effect of $\theta_A$ assuming that $\mu$ is real
and positive. 
Variations of the EDMs in the ($M_2$,$m_0$) plane are displayed in 
Fig.~\ref{fig:cont-m0-m2} for 
$\tan\beta=10$, $m_A=700\,{\rm GeV}$, 
$|\mu|=500\,{\rm GeV}$, $|A|=500\,{\rm GeV}$, 
$\theta_\mu=0$, $\theta_A=\pi/6$. 
The result for the neutron EDM is shown in the upper left window.
If we assume previous experimental limit $d_n$ $\lesssim$ $10^{-25}$
using the neutron itself, 
only the small mass region $m_0$, $M_2$ $\lesssim$ $1\,{\rm TeV}$ 
would be excluded. However, with the new limit
\begin{eqnarray}
0.06\times 10^{-28}{\rm e\cdot cm}<d_n<0.68\times 10^{-28}{\rm e\cdot cm}
\label{eq:dn_exp}
\end{eqnarray}
derived from eq.(1.3), 
it is found that the large region below the solid line is excluded 
by the neutron EDM.

In the upper right window of the same figure, we plot the contours
of the proton EDM for the same choice of the parameters. 
In this case, 
$d_p$ is negative in the whole region. As for the magnitude, 
$|d_p|$ is of the same order of magnitude as $d_n$. 
This can be seen more clearly in the contour plot of the ratio
$d_p/d_n$ (lower left window). In the region $M_2 \gg m_0$, 
$|d_p|$ is smaller than $d_n$, while it is nearly equal to 
$d_n$ for  $M_2 \lesssim m_0$. 

For comparison, we also plot the contours of $d_e$ in the 
lower right window. 
The current experimental upper bound for the electron EDM is
$|d_e| < 1.6\times 10^{-27}{\rm e\cdot cm}$ \cite{reg}. 
We can see that the constraint from $d_n$ is much severer than
that from $d_e$.

\begin{figure}[p]
\begin{center}
\begin{minipage}{9cm}
\epsfig{file=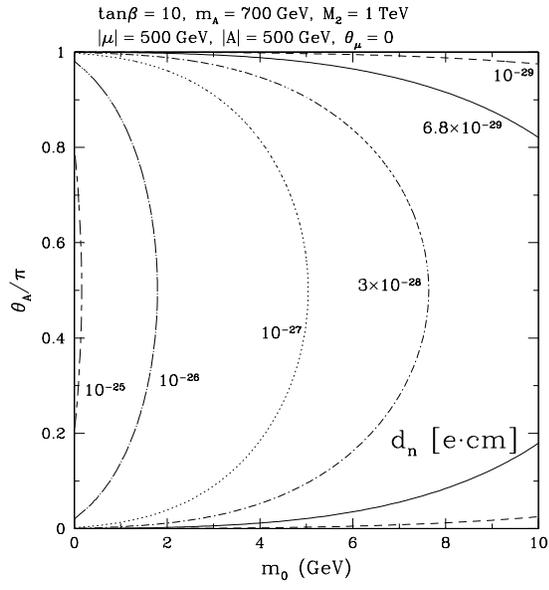,width=8cm} 
\end{minipage}
\end{center}
\caption{\label{fig:cont-tha0-m0} 
Contours of $d_n$ in the ($m_0$,$\theta_A$) plane
for $\tan\beta=10$, $m_A=700\,{\rm GeV}$, $M_2=1\,{\rm TeV}$, 
$|\mu|=500\,{\rm GeV}$, $|A|=500\,{\rm GeV}$, $\theta_\mu=0$. 
}
\end{figure}

The effect of varing $\theta_A$ can be seen in the ($m_0$,$\theta_A$) plane
in Fig.~\ref{fig:cont-tha0-m0}
for $\tan\beta=10$, $m_A=700\,{\rm GeV}$, $M_2=1\,{\rm TeV}$, 
$|\mu|=500\,{\rm GeV}$, $|A|=500\,{\rm GeV}$, $\theta_\mu=0$. 
For $0<\theta_A<\pi$, $d_n$ is positive, and only small regions outside the
solid lines are allowed by the neutron EDM constraint (\ref{eq:dn_exp}). 
In particular, for $0.2\pi \lesssim \theta_A \lesssim 0.8\pi$, 
the common sfermion mass $m_0$ must be larger than 10 TeV 
to satisfy (\ref{eq:dn_exp}) for this choice of the parameters. 
For $\pi<\theta_A<2\pi$, $d_n$ is negative.

\begin{figure}[p]
\begin{center}
\epsfig{file=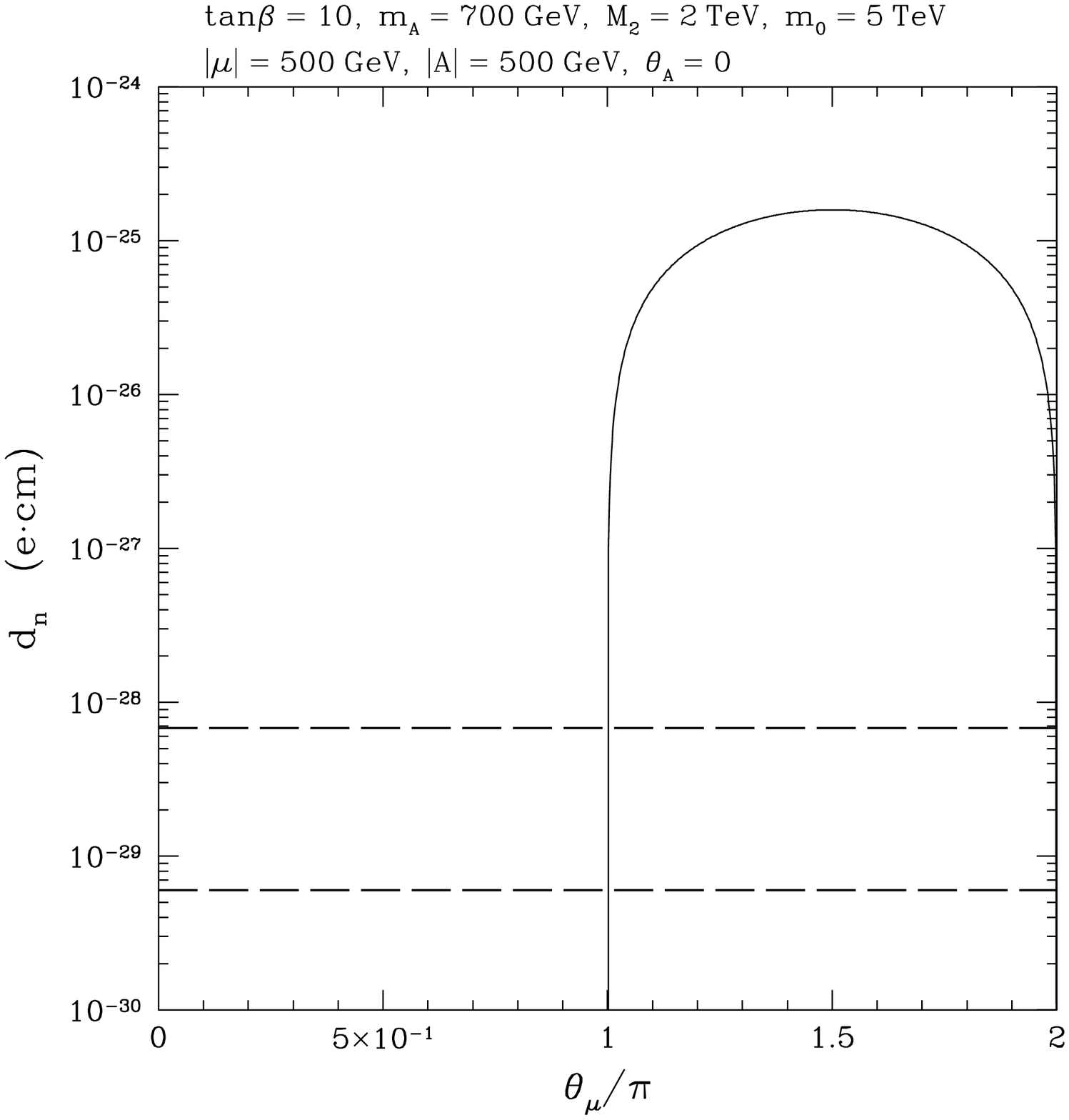,width=8cm} \\
\epsfig{file=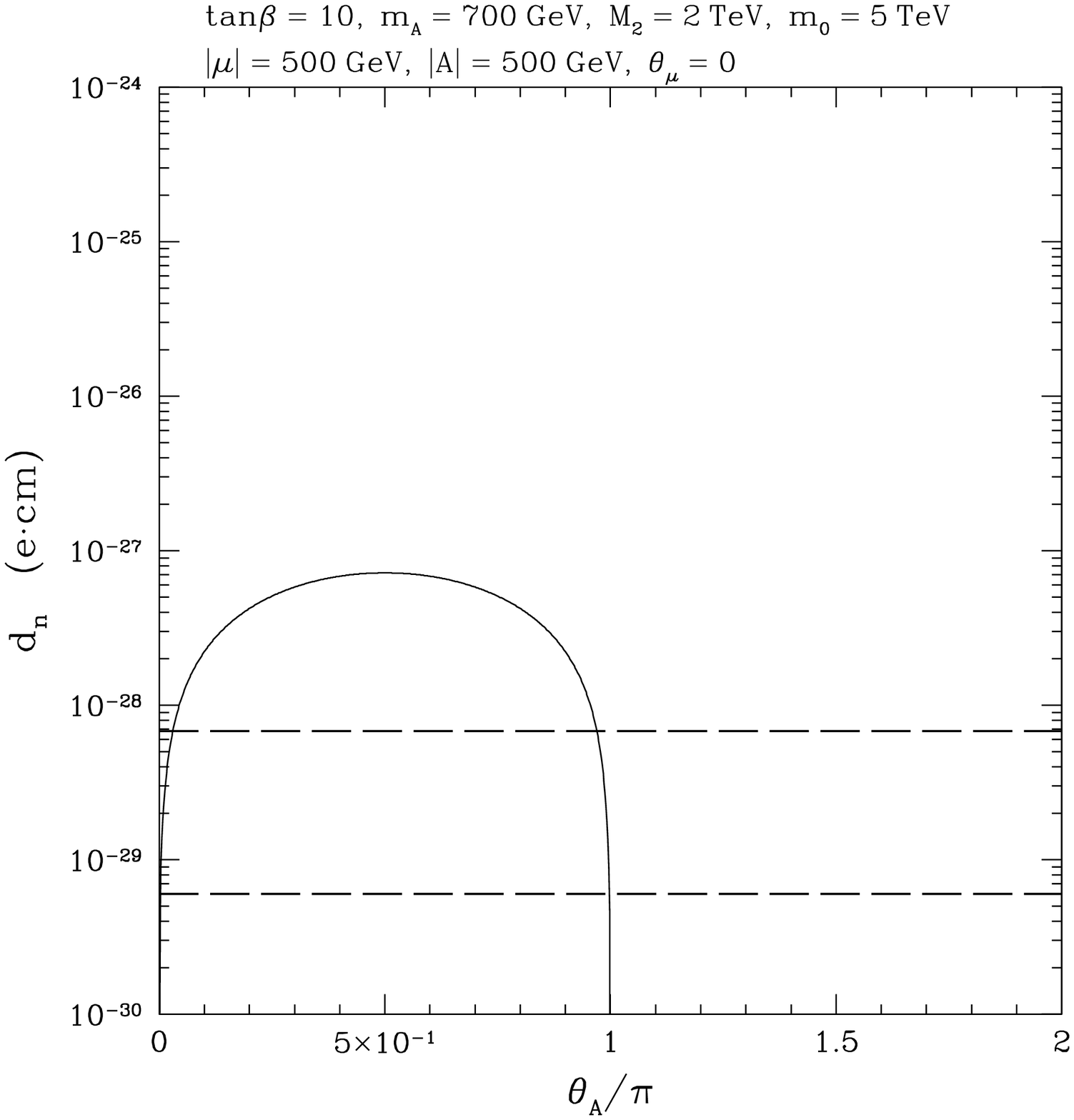,width=8cm}
\end{center}
\caption{\label{fig:dn-theta} 
The neutron EDM $d_n$ vs $\theta_{\mu}$ (upper window) 
and $\theta_A$ (lower window). 
Relevant supersymmetric parameters for the upper window are chosen as 
$\tan\beta=10$, $m_A=700\,{\rm GeV}$, 
$M_2=2\,{\rm TeV}$, $m_0=5\,{\rm TeV}$, 
$|\mu|=500\,{\rm GeV}$, $|A|=500\,{\rm GeV}$ and $\theta_A=0$,
while those for the lower window are the same except $\theta_\mu=0$. 
The range $0<\theta_{\mu}<\pi$ in the upper window and
$\pi<\theta_A<2\pi$ in the lower window corresponds to $d_n<0$. 
In the region between the two dashed lines, the EDM constraint 
$0.06\times 10^{-28}{\rm e\cdot cm}<d_n<0.68\times 10^{-28}{\rm e\cdot cm}$
is satisfied. 
}
\end{figure}

Finally, we discuss the effect of $\theta_\mu$. 
The phase dependence of the neutron EDM for fixed $m_0$ and $M_2$
is shown in Fig.~\ref{fig:dn-theta}. 
The relevant supersymmetric parameters for the upper window are chosen as 
$\tan\beta=10$, $m_A=700\,{\rm GeV}$, $M_2=2\,{\rm TeV}$, $m_0=5\,{\rm TeV}$, 
$|\mu|=500\,{\rm GeV}$, $|A|=500\,{\rm GeV}$ and $\theta_A=0$,
while those for the lower window are the same except $\theta_\mu=0$. 
The range $0<\theta_{\mu}<\pi$ in the upper window and the range 
$\pi<\theta_A<2\pi$ in the lower window corresponds to $d_n<0$. 
In the region between the two dashed lines, the EDM constraint 
(\ref{eq:dn_exp}) is satisfied. 
From the two plots, we see that the CP-violating phases $\theta_{\mu}$
and $\theta_A$ are severely bound to zero even in a multi-TeV region 
of SUSY breaking mass parameters. Relatively speaking, $\theta_A$ 
is less constrained than $\theta_{\mu}$. 
Therefore, if we take sizable $\theta_{\mu}$ rather than $\theta_A$,
the constraints on $m_0$ and $M_2$ become severer than those in 
figures \ref{fig:cont-m0-m2} and \ref{fig:cont-tha0-m0}.

\vspace{0.5cm} 
\section{Discussions 
on $d_e $ and $d_p $ } 

The expected magnitudes of the electron EDM $d_e$ and the proton 
EDM $d_p$ depend on the parameters of the SUSY models. It should be 
rather difficult to reliably predict the relative magnitude between $d_e$ and $d_p$. 

Therefore, the contributions from $d_e$ and $d_p$ to the atomic EDM values 
should be estimated carefully. In this paper, we do not include any contributions 
of the electron EDM, but instead we will discuss the electron EDM 
contribution to the atomic EDM in a qualitative fashion. 

Below, we evaluate the contributions from $d_e$ and $d_p$ to the Nitrogen and 
Thallium atomic EDMs. 

\subsection{Electron EDM in atomic system} 

The electron EDM contributions to the atomic EDM come from the relativistic 
effects of the electron wave functions. This estimation is carefully done 
in \cite{q44}, and the atomic EDM $d_Z^{\rm atom}$ can be written as
$$ d_Z^{\rm atom} \simeq C_1 (Z\alpha)^2 \left(1+2(Z\alpha)^2 \right) d_e . 
\eqno{(4.1)} $$
If we take $C_1 \sim -100$, then we obtain for Nitrogen and T$\ell$ cases, 
$$ d_{N}  \simeq -0.3 d_e \eqno{(4.2a)} $$
$$ d_{T\ell}  \simeq -60 d_e \eqno{(4.2b)} $$
where the sign in front of the number is not determined from eq.(4.1) since 
it depends on each atomic state. 

The elaborate calculation for T$\ell$ case shows 
$$ d_{T\ell} \simeq -585 d_e . \eqno{(4.2c)} $$
The order of magnitude enhancement may come from the atomic state configuration. 

\subsection{Proton EDM in atomic system} 

As we discussed in section 2, the proton EDM mainly comes from the nuclear excitations. 
The proton EDM contribution to the atomic EDM can be written as
$$ d_Z \simeq  0.03 Z d_p . \eqno{(4.3)}  $$
In this case, we obtain for Nitrogen and T$\ell$ cases, 
$$ d_{N} \simeq -0.2 d_p \eqno{(4.4a)} $$
$$ d_{T\ell} \simeq 2.4 d_p . \eqno{(4.4b)} $$
On the other hand, as we saw in section 2, the elaborate calculations show 
$$ d_{N} \simeq -0.16 d_p \eqno{(4.4c)} $$
$$ d_{T\ell} \simeq 4.8 d_p . \eqno{(4.4d)} $$
As can be seen from eqs.(4.4), the qualitative formula of eq.(4.3) gives 
sufficiently reliable estimations of the nuclear EDM. 

\subsection{$d_e$ and $d_p$ from SUSY models } 

Now, as we saw in the previous section, the relation between 
the $d_e$ and $d_p$ EDMs from the SUSY model calculations are shown in fig. 1. 
If we choose the values of $M_2$ and $m_0$ to be 
$$ M_2 \simeq 9 \ \ {\rm TeV}, \ \ \ \ \ m_0 \simeq 9 \ \ {\rm TeV} $$
then we obtain
$$ d_e \simeq -2.8\times 10^{-31} \  {\rm e}\cdot {\rm cm}  \eqno{(4.5a)} $$ 
$$ d_p \simeq -1.9\times 10^{-29} \  {\rm e}\cdot {\rm cm} . \eqno{(4.5b)} $$ 
In this case, the atomic EDM for Nitrogen and T$\ell$ cases from the electron EDM  
become 
$$ d_{N}^e \simeq -1.0\times 10^{-31} \  {\rm e}\cdot {\rm cm}  \eqno{(4.6a)} $$ 
$$ d_{T\ell}^e \simeq -(0.2 \sim 1.8 )\times 10^{-28} \  {\rm e}\cdot {\rm cm} . 
\eqno{(4.6b)} $$ 
On the other hand, we obtain the atomic EDM for Nitrogen and T$\ell$ cases 
from the proton EDM contribution as
$$ d_{N}^p \simeq -0.4\times 10^{-29} \  {\rm e}\cdot {\rm cm} . \eqno{(4.6c)} $$ 
$$ d_{T\ell}^p \simeq -1.0\times 10^{-28} \  {\rm e}\cdot {\rm cm} . \eqno{(4.6d)} $$ 
Therefore, the EDM contributions from the electron (relativistic effects) and 
the proton (nuclear finite size effects) for the T$\ell$ case 
may well be comparable. But the atomic EDM for the Nitrogen case is mainly 
determined from the proton EDM.

\vspace{0.5cm} 
\section{Conclusions} 
We have presented the proton and neutron EDMs which are extracted from 
the atomic EDM measurements. In particular, we show that the proton EDM 
can be obtained if one can measure the EDM of Nitrogen atomic systems. 
The EDM measurement of Nitrogen atomic system enables to discuss the flavor 
dependence of the quark EDM which should give some important information 
on the origin of the T-violation in field theory models.

Also, we have studied the EDM of neutron, proton and electron in 
the minimal supersymmetric standard model. 
In the presence of supersymmetric CP-violation, 
the new limit on the neutron EDM extracted from atomic systems 
excludes a wide parameter region of SUSY breaking masses above 1 TeV,
while the old limit excludes only a small mass region below 1 TeV. 
Thus the observed neutron EDM is found to give a very stringent test on the 
supersymmetric model.

\vspace{1cm}


\end{document}